\documentclass[aps,twocolumn,superscriptaddress,showpacs]{revtex4}
\usepackage{graphicx}
\usepackage{amsmath}
\newcommand{\beq}{\begin{equation}}
\newcommand{\eeq}{\end{equation}}
\newcommand{\bea}{\begin{eqnarray}}
\newcommand{\eea}{\end{eqnarray}}
\newcommand{\pdag}{{\phantom{\dagger}}}

\begin{document}
\title{Read out of a Nuclear Spin Qubit   }
\author{M.\ Kindermann}
\affiliation{ Department of Physics, Massachusetts Institute of Technology,
Cambridge MA 02139, USA}
\author{D.\ G.\ Cory}
\affiliation{ Department of Nuclear Engineering, Massachusetts Institute of Technology,
Cambridge MA 02139, USA}
\date{May 2004}
\begin{abstract}
 We propose  a detector to read out  the state of  a single nuclear spin, with
 potential application in future  scalable NMR quantum computers. It is based on
 a ``spin valve'' between bulk nuclear spin systems that is highly sensitive to
 the state of the measured spin. We suggest a concrete realization of that
 detector in a Silicon lattice. Transport of spin through the proposed spin valve is analogous to that of charge through an electronic nanostructure, but exhibits distinctive new features.   
\end{abstract}
\maketitle

Nuclear magnetic resonance (NMR) experiments have been a valuable testbed for quantum information processing (QIP)  \cite{Cor00} and they still provide the largest collections of coupled qubits available at present \cite{Kni00,Van01}. Most NMR  QIP experiments are  performed on liquids  and  suffer from the lack of  scalability.   Solid state spin systems have been proposed as a promising route to  scalability in NMR QIP \cite{Kan98,Sut02,Lad02}. Their experimental implementation is, however, challenging. One major obstacle that has to be overcome is    the read out problem.   In most proposals it requires the measurement of the quantum state of a single nuclear spin. Experiments that have successfully detected single spin resonances are promising \cite{Koe93,Wra93,Wra97,Rug04,Xia04}. Recently the read out of a single electron spin has been reported \cite{Elz04}. The measurement of the state of a single nuclear spin has  remained, however, elusive up to now. The adiabatic transfer of the spin state of a  nucleus to that of an electron \cite{Kan98} or magnetic resonance force microscopy \cite{Ber01} have been proposed to read out an NMR qubit. These techniques introduce, however, unwanted sources of additional decoherence \cite{Fu03}.  The use of ensembles of nuclear spins in $^{29}{\rm Si}$ as qubits has been proposed \cite{Lad02} to enhance the measurement signal.   Optical detection by measuring the energy of photons emitted by bound excitons is another proposal to measure a single nuclear spin \cite{Fu03}. Most of the above mentioned  schemes require very specific  material properties for their implementation. In this Letter we propose a scheme   for the read out of a nuclear spin qubit that relies only on the dipolar coupling between nuclear spins. That interaction is generically present in solids.  
       
\begin{figure}
\includegraphics[width=7cm]{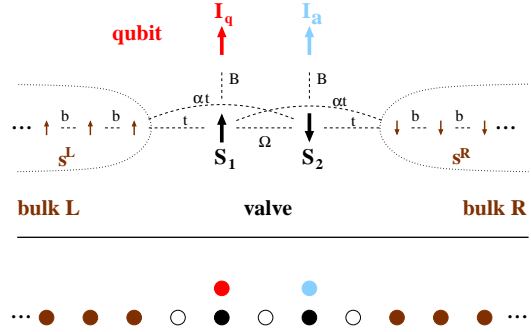}
\caption{ Proposed nuclear spin detector. Upper panel: schematic setup with inter-spin coupling constants. Lower panel: Possible realization in a  Silicon lattice. Filled circles: $^{29}{\rm Si}$. Open circles: $^{28}{\rm Si}$.    }  \label{fig1} 
\end{figure}

Our proposal is inspired by charge detectors that utilize quantum point contacts (QPC) \cite{QPC1,QPC2,QPC3,QPC4,QPC5,QPC6}. Such a detector has recently been successfully employed for the read out of an electron spin qubit after spin to charge  conversion \cite{Elz04}. Converting a nuclear spin into a charge signal is hard and we therefore propose to directly measure a nuclear spin by an analogue of a QPC for nuclear spin currents, as shown in Fig.\ \ref{fig1}. Our detector consists  of two bulk  systems of nuclear spins ${\bf s}^{L/R}_j$   connected by two spins ${\bf S}_j$ that act as a spin valve.  A difference in polarization of the bulk spins to the two sides of the valve can drive an equilibrating spin current between them. A qubit spin ${\bf I}_q$ and an auxiliary spin ${\bf I}_a$ (that is prepared in a known state) create local magnetic fields for ${\bf S}_1$ and ${\bf S}_2$. These fields depend on the qubit's state and can bring the valve spins into and out of resonance.    We demonstrate below that as a consequence the spin current between the bulk systems is highly sensitive to  the state of the qubit.    After an appropriate measurement time the state of the qubit  is therefore encoded in the spin state of a measurable number (typically $10^6$) of  bulk spins and can be read out  with standard techniques. 

We write the Hamiltonian of  the spin detector  Fig.\ \ref{fig1} as   the sum
\beq
H=H_b+H_v+H_T.
\eeq
 We assume  the bulk spins $ {\bf s}^{L/R}_j$ to form a linear chain of $2N$ sites ($N$ is large) with Hamiltonian
\beq
 H_b=b\sum_{j=0}^{N}{\left( s_j^{Lx}s_{j+1}^{Lx} +s_j^{Ly}s_{j+1}^{Ly}+ \{L\leftrightarrow R\}\right)} .
 \eeq
 $H_v$ describes the valve spins ${\bf S}_j$  coupled to the qubit  ${\bf I}_q$, 
  \beq
 H_v= 2\Omega(S^x_1S^x_2+S_1^yS_2^y) +B(I^z_q S^z_1+I^z_a S^z_2),
 \eeq
 and $H_T$ couples the valve spins to the bulk,
 \bea
 H_T&=& 2 t[s^{Lx}_0(S^x_1+\alpha S^x_2)+s^{Ly}_0(S^y_1+\alpha S^y_2)]\nonumber \\
 &+&2 t[s^{Rx}_0(S^x_2+\alpha S^x_1)+s^{Ry}_0(S^y_2+\alpha S^y_1)].
\eea
The choice of coupling of the qubit to the spin valve  $B I^z_q S^z_1$ is crucial. It assures that the qubit's state  is conserved after  having been projected onto the logical basis $\{|\!\! \uparrow\rangle,|\!\!\downarrow\rangle\}$   ($ I^z_q |\!\!\uparrow\rangle= |\!\!\uparrow\rangle/2 $ and $ I^z_q |\!\!\downarrow\rangle=- |\!\!\downarrow\rangle/2$). This allows the detector to be operated until it has accumulated a detectable change in bulk  polarization.  Experimentally this form of the coupling Hamiltonian can be implemented by choosing the detector nuclei to be of a sort different from that of the qubit nucleus. A difference of the gyromagnetic ratios of the two kinds of nuclei  together with a strong magnetic field along the z-direction then eliminates the couplings $I^x_q S^x_1$ and $I^y_q S^y_1$ that are otherwise present. It removes the corresponding transitions far off resonance and leads to an effective Hamiltonian of the form  $H_v$.
 By means of a Jordan-Wigner transformation the spin operators in $H$ can be expressed in terms of fermion operators $c^{L/R}_j$ and $d_j$ with standard anti-commutation relations \cite{Fel98}.
In Fourier space, $c^{\eta}(p)=(1/\sqrt{N})\sum_j c^{\eta}_j\exp ipj$ ($\eta\in\{L,R\}$), the Hamiltonian takes the form
\bea
H_b&=&b\sum_
{\eta\in\{L,R\},p}\cos p\,c^{{\eta}\dag}(p)c^{\eta}(p) \nonumber \\
H_v&=& \Omega(d_1^{\dag}d^{\pdag}_2+h.c.)+B[I^z_q(d^{\dag}_1 d^{\pdag}_1-{\textstyle \frac{1}{2}})+I^z_a(d^{\dag}_2 d^{\pdag}_2-{\textstyle \frac{1}{2}})] \nonumber \\
H_T&=& \sum_
{\eta\in\{L,R\}} c^{\eta\dag}_0 T_{\eta}+h.c..
\eea
 Both  $I_q$ and $I_a$ are  not dynamical and $H_b$ and $H_v$ are readily diagonalized in their fermionic form. The non-linear terms
\bea \label{Teta}
T_L&=&t(d_1+\alpha e^{i\pi d^{\dag}_1d^{\pdag}_1} d_2), \nonumber \\
T_R&=&t(d_2+\alpha e^{-i\pi d^{\dag}_2d^{\pdag}_2} d_1),
\eea
however, introduce an interaction between the fermions. This precludes a straightforward analytic solution of the transport problem.
As we show below, these interaction terms affect the transport behavior of the spin valve in a qualitative way.   In Ref.\ \cite{Mic04}   similar interactions have been encountered for hard core bosons. There they were dealt with numerically. Here we exploit the fact, that in typical NMR experiments the temperature $T$ is  high compared to all intrinsic energy scales.  This  allows for a  modified mean field treatment of the bulk spins that can be carried out analytically.  In this limit, the bulk spins rearrange to their equilibrium state instantly after every transfer of spin into the bulk. To a good approximation their dynamics is therefore independent of that of the valve spins.    This reduces the complexity of the problem to the four dimensional Hilbert space of the two valve spins ${\bf S}_j$. It leads to a quantum master equation for the reduced density matrix of the valve spins. Such an approach  often proves useful in problems of quantum transport \cite{Gur96}. It is valid if the time scale $\tau_c=1/\max\{\Omega, B, t^2/b\}$ of the dynamics of ${\bf S}_j$ is much longer than that of the bulk dynamics $\tau_b=1/\min\{T,b\}$, $\tau_c \gg \tau_b$.

In deriving this master equation we proceed closely along the lines of Ref.\ \cite{art12}. We only repeat the main steps here.  We evaluate the generating function   
\beq  \label{Z}
 {\cal Z}(\lambda)={\rm Tr}\, e^{-i \lambda_{\eta} S^{\eta} }\, e^{-iH\tau}\, e^{ i\lambda_{\eta} S^{\eta} } \, \rho^{\rm (in)}\, e^{iH\tau}.
 \eeq
 $\rho^{\rm (in)}$ is the initial density matrix of the spin system and   $S^{L/R}=\sum_j s_j^{L/Rz}$ is the total bulk spin.
  ${\cal Z}$  generates moments of the amount $\Delta S^{\mu}$ of spin that is transferred through the valve spins during time $\tau$. In particular, the mean amount of transferred spin averaged over identical experiments is 
  \begin{equation} \label{genZ}
 \left\langle  \Delta S^{\eta}  \right\rangle =  i \frac{d}{d \lambda_{\eta}}   \ln {\cal Z}(\lambda)\Big|_{\lambda=0}.
 \end{equation}
It is crucial to note, that the Hamiltonians $H$ in Eq.\ (\ref{Z}) as well as $ S^{\eta}$ are quadratic in the bulk fermion operators $c_i^{\eta}$. Hence they can be integrated out exactly,  yielding  
 \bea \label{Zla}
 {\cal Z}(\lambda) &=&   {\rm Tr}_c\,  {\cal T}_{\pm} \,  \rho^{\rm (in)}_c\, e^{ -i\int_0^{\tau}{dtdt'\, \sum_{\eta\in\{L,R\}}\tilde{T}_{\eta}^{\dag}(t) G_{\eta}(t-t')\tilde{T}_{\eta}(t') }} \nonumber \\
 &&\!\!\!\!\! \times e^{-i\int_0^{\tau}{dt\, \Omega  (d_1^{\dag}\tau^z d^{\pdag}_2+h.c.)+B(I^z_q\, d^{\dag}_1 \tau^z d^{\pdag}_1+I^z_a\, d^{\dag}_2 \tau^z d^{\pdag}_2})} .
  \eea
Here, all operators are vectors in a ``Keldysh space'' $(d_j^+,d_j^-)$ of operators $d_j^+$ and $d_j^-$ that originate  from the first and   the second exponential of Eq.\ (\ref{Z}) respectively. By the symbol ${\cal T}_{\pm}$ they are ordered in time as well as relative to the initial reduced density matrix $\rho^{\rm (in)}_c$ of the central spins ${\bf S}_j$ and  ${\bf I}_q/{\bf I}_a$. $\tau^z$ is  the third Pauli matrix  and $\tilde{T}_{\eta}=\tau^z\exp(-i\lambda_{\eta} \tau^z) T_{\eta}$.  
The ``mean field`` due to the bulk spins is able to increase or decrease ${\bf S}_j$  and it is quantified   by the bulk spins' Green functions $G_{\eta}$. $G_{\eta}(t)$  are peaked around $t=0$ with width $\tau_b$. In the high temperature regime $\tau_c \gg \tau_b$ of interest here the exponent in Eq.\ (\ref{Zla}) is therefore local in time and ${\cal Z}$ is the integral of an ordinary differential equation. We write it accordingly as the trace over a time-dependent density matrix,
\beq
 {\cal Z}(\lambda) =  {\rm Tr}_c\,  \rho^{\lambda}_c(\tau), 
 \eeq
 that obeys the master equation
 \bea \label{qm}
 \partial_t  \rho^{\lambda}_c & = & {\cal L}[ \rho^{\lambda}_c] = -i \left[H_v+p_{\eta} (T_{\eta}^{\dag}T_{\eta}^{\pdag}+
 T_{\eta}^{\pdag} T_{\eta}^{\dag}),\rho^{\lambda}_c\right] \nonumber \\
 & &\mbox{} - \frac{1}{2b} \left[ n_{\eta} T_{\eta}^{\pdag}T_{\eta}^{\dag} \rho^{\lambda}_c + (1-n_{\eta}) T_{\eta}^{\dag}T_{\eta}^{\pdag} \rho^{\lambda}_c +h.c.\right] \nonumber \\
 &&\mbox{} +\frac{1}{b}\left[ n_{\eta}  T_{\eta}^{\dag} \rho^{\lambda}_c T_{\eta}^{\pdag} \,e^{i\lambda_{\eta}}+(1-n_{\eta}) T_{\eta}^{\pdag} \rho^{\lambda}_c T_{\eta}^{\dag}\,e^{-i \lambda_{\eta}} \right] \nonumber\\
  \eea
with  initial condition $ \rho^{\lambda}_c(0)= \rho^{\rm in}_c$. 
Due to the coupling to  the bulk spins,   $ \rho^{\lambda}_c$ is  evolved in time by a linear ``superoperator'' ${\cal L}$ of Lindbladian form \cite{Pre} rather than by a Hamiltonian.  $n_{\eta}-1/2={\rm Tr}\, \rho^{\rm in} s_0^{\eta\,z} $ are the expectation values of the z-projections of the bulk spins and  $p_{\eta}= (2\pi b)^{-1}{\cal P}\int{d\omega\,n_{\eta}/\omega}$ are principal value integrals that are cut-off at high frequencies by the bandwidth $b$ of the bulk spin excitations.   We compute ${\cal Z}$ at large times $\tau$ by exponentiating the largest eigenvalue of $\tau {\cal L}$.  The spin currents $j^{\eta}=\Delta S^{\eta}/\tau$ then follow from Eq.\ (\ref{genZ}).   We take an initial density matrix $\rho^{\rm in}_c$  corresponding to $I_a$ being prepared in state $|\!\!\uparrow\rangle$. If the qubit's state is  $|\!\!\downarrow\rangle$, the  spin current  has magnitude
\begin{widetext}
\beq \label{ME}
j^R_{\downarrow}=2 (1+\alpha^2) \Delta n\Gamma \frac{\alpha^2 B^2 +(1-\alpha^2)^2[\alpha^2\Gamma^2(1-\xi \Delta n^2)-2\xi \alpha \tilde{\Omega} B \Delta n(1-\alpha^2)^{-1}+\tilde{\Omega}^2]}{(1+\alpha^2)^2 B^2 + (1-\alpha^2)^2[(1+\alpha^2)^2\Gamma^2 - 4 \xi \alpha \tilde{\Omega} B \Delta n(1-\alpha^2)^{-1}+4\tilde{\Omega}^2]}.
\eeq
\end{widetext}
 Here, $\Gamma=t^2/b$, $\tilde{\Omega}=\Omega+4\alpha t^2(p_L+p_R)$,  $\Delta n= n_L-n_R$,  and $\xi=1$. If on the other hand $I_q$ is in  state  $|\!\!\uparrow\rangle$, the magnitude of the  spin current is $  j^R_{\uparrow}=j^R_{\downarrow}|_{B=0}$.  The measurement contrast $C$, defined as the ratio $C=j^R_{\uparrow}/j^R_{\downarrow}$ of the signals for both possible states of $I_q$, characterizes the performance of the detector \cite{footnote}. 
 \begin{figure}
\includegraphics[width=7cm]{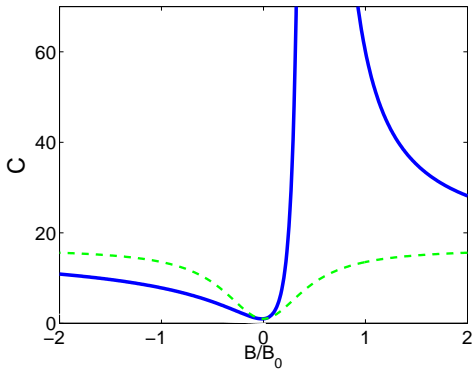}
\caption{Measurement contrast $C$ of our detector as a function of the strength $B$ of its coupling to the qubit - normalized to the coupling $B_0$ to a proton qubit (solid line). Dashed line: $C$ for a corresponding system of free fermions.  }  \label{fig2} 
\end{figure}

Fig.\ \ref{fig1} suggests an implementation of our proposal  in a chain of Silicon atoms \cite{Ito03}. The weak links between the bulk and the valve spins are realized by lattice vacancies or isotopes $^28$Si without nuclear spin. In typical solids, nuclear spins are coupled by the dipolar interaction that falls off as $r^{-3}$ with the distance $r$ between spins. For  single vacancies we therefore have $t=\Omega=b/8$ and $\alpha=1/8$.  
 In Fig.\ \ref{fig2} we plot the measurement contrast $C$ for these parameter values at full polarization ($n_L=1$, $n_R=0$) as a function of the coupling strength $B$. $B$ is normalized to  the value $B_0$ that one has for the measurement of a proton spin. 
 We estimate $B_0=(8/5)^3(b/4)$.  Fig.\ \ref{fig2} clearly demonstrates that our detector yields good contrast over a large range of coupling strengths. In particular, we find excellent contrast $C\approx 60$ for the measurement of a proton spin.

\begin{figure}
\includegraphics[width=7cm]{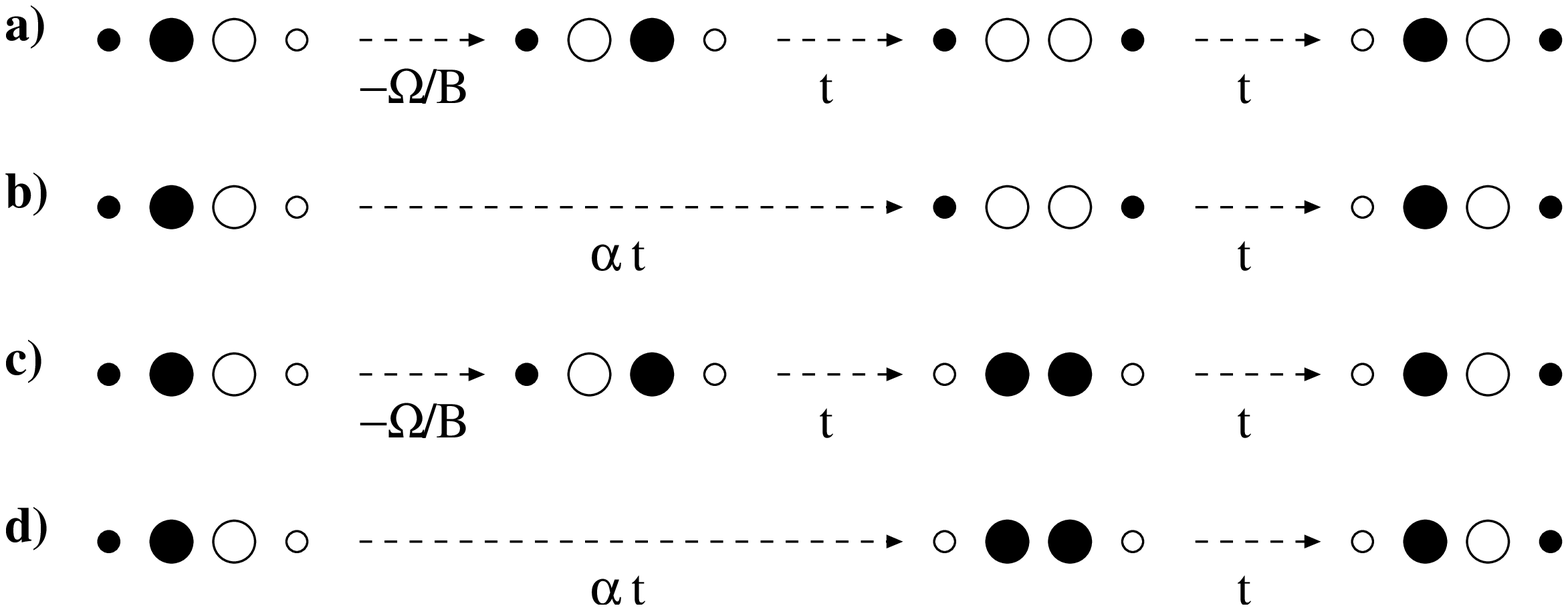}
\caption{ Transport processes for a typical initial spin configuration to lowest order in perturbation theory. Large circles: spins $S_1$ and $S_2$. Small circles: bulk spins $s^L_0$ and $s^R_0$. Filled circles symbolize spin up (occupied site), empty circles  spin down (unoccupied site) for a spin (fermion) system.   }  \label{fig3} 
\end{figure}

   For  full polarization  the measurement contrast can in fact be made
   arbitrarily large by fine tuning $B$. This important property of our detector
   can be traced to the presence of the interaction terms in Eqs.\ (\ref{Teta}).
   This follows from a comparison of $C$ for our spin system with that of the
   free electron system described by $H$ without the phases $\exp(i\pi
   d^{\dag}_jd^{\pdag}_j)$ in $T_{\eta}$ (that descibes a double quantum dot
   \cite{Gur96}).  $C$ in that case is obtained from Eq.\ (\ref{ME}) by setting $\xi=0$ and it is bounded from above, as shown in Fig.\ \ref{fig2}. 
 The discussed divergence of $C$ for the spin valve  is due to an interference effect that causes $ j^R_{\downarrow}$  to vanish at 
\beq
B_{int}=(1-\alpha^2) \frac{\tilde{\Omega}}{\alpha}.
\eeq
At this coupling strength, there occurs a complete destructive interference
of transport processes involving next-nearest neighbor couplings $\alpha t$ with processes involving nearest neighbor couplings $t$ and $\Omega$ only. To illustrate this, we show  in Fig.\ \ref{fig3} all lowest order transport processes for a typical initial spin configuration. Adding up their amplitudes leads to a transport rate $\propto |t^2(\Omega/B-\alpha)|^2$ that indeed vanishes at $B=B_{int}$ (to lowest order in $\alpha$ and $t$). For free fermions the same processes exist with the correspondences spin-up $\leftrightarrow$ occupied site, spin-down $\leftrightarrow$ unoccupied site. However, the last process d) acquires an additional minus sign, because the fermions in the final state are interchanged with respect to those in the final state of  process b). This results in a strictly non-vanishing rate $\propto |t^2\Omega /B|^2$ that is symmetric under the reversal of the sign of $B$. These features carry over to all orders of perturbation theory and render transport through the spin valve qualitatively different from that through the corresponding system of free fermions, as seen clearly in Fig.\ \ref{fig2}.

\begin{figure}
\includegraphics[width=7cm]{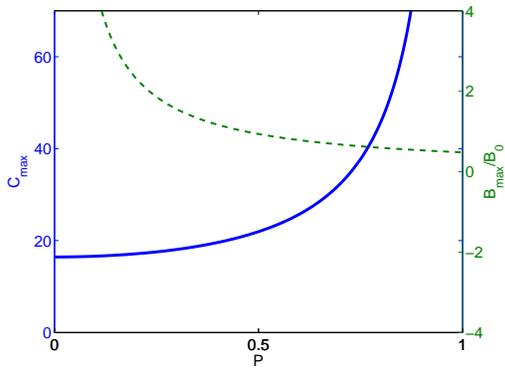}
\caption{Maximal contrast $C_{max}$ (solid line) together with the the optimal coupling strength $B_{max}$ at which it is attained (dashed line) as a function of bulk polarization $P$. }  \label{fig4} 
\end{figure}

 In case the polarization of the bulk spins  is not  complete, $C$ does not diverge anymore. Instead, it possesses a maximum at an optimal coupling strength $B_{max}$.  We plot $C_{max}$ and $B_{max}$ as functions of the polarization $P$  in Fig.\ \ref{fig4}, taking $n_L= (1+P)/2$ and $n_R=(1-P)/2$.   Evidently good contrast can be attained even for a  weak polarization of the bulk spins. Note, though, that the plotted maximal contrast is in practice increasingly hard to achieve for decreasing polarization, since the necessary optimal coupling $B_{max}$ diverges at $P=0$.
 
We finally comment on the applicability of our model to real spin systems. We have analyzed a one-dimensional spin chain and have neglected next-nearest neighbor couplings as well as $s^{\eta z}_j s^{\eta z}_{j+1}$ interaction terms for the bulk spin systems. In principle, effectively one-dimensional spin chains are available in fluorapatite \cite{Cho96} and proposed in Silicon by Itoh. The mentioned couplings can be eliminated with experimental effort \cite{Ram04}.   We believe, however, that our predictions remain qualitatively correct also in more general situations. Spin systems with $s^{\eta z}_j s^{\eta z}_{j+1}$ interaction are described by interacting fermions  \cite{Fradkin}. To lowest order in $t$ our Eq.\ (\ref{Zla}) holds also in the presence of   bulk  fermion interactions, with $G_{\eta}$ being the Green function of the interacting system.  For one-dimensional interacting fermions $G_{\eta}$ exhibits an anomalous  behavior that is cut-off at low energies by the temperature of the system. In our limit of  temperatures that are larger than the bandwidth we do therefore not expect that these interactions have a  qualitative effect. The same reasoning applies to possible  next-nearest neighbor couplings and higher dimensionality, that are similarly  described by additional interactions between bulk fermions in our model.  Inclusion of the direct dipolar coupling between the spins $s^{L}_0$ and $s^R_0$ leads to a leakage current that is suppressed by a factor of  $3^{-6}\approx 10^{-3}$ relative to the leading contribution. The coupling $S^z_1 S^z_2$ has no effect on spin transport and to lowest order in $t$ the couplings  $s^{\eta z} S^z_j$  result in a trivial shift of the energies of $S^z_j$.
 
 In conclusion, we have proposed a local read out scheme for future solid state NMR quantum computers. It is based on a spin valve between bulk nuclear spin systems. Our analytical results show that it is highly sensitive to the state of a  nuclear spin qubit. It  implements  a detector  of single nuclear spins. An experimental realization of the proposed detector would not only solve a major problem on the way to a scalable NMR quantum computer. Our comparison with the corresponding electronic system indicates that it would also provide us with a system that exhibits novel and interesting phenomena in quantum transport.
 
 We acknowledge financial support by CMI.


\begin{thebibliography}{99}
\bibitem{footnote}
  Measurement errors due to  statistical fluctuations of the transferred polarization  can be obtained from ${\cal Z}$ by taking higher derivatives.  For measurable polarization transfer  $\Delta S^R \approx 10^6$ they are suppressed by  factors of $1/\sqrt{|\Delta S^R|}\approx 10^{-3}$ and are not of practical importance.
\end{thebibliography}
\end{document}